\documentclass[11pt,a4paper]{article}
\usepackage{mathrsfs}
\usepackage{amsfonts}
\usepackage{jheppub}

\usepackage{amsmath,graphicx,amsfonts, mathrsfs, amssymb}
\usepackage{amsmath,epsfig}
\usepackage{amssymb,amsfonts}
\usepackage{latexsym}
\usepackage{epsfig}
\newbox\pippobox
\def\be{\begin{equation}}
\def\ee{\end{equation}}
\def\bea{\begin{eqnarray}}
\def\eea{\end{eqnarray}}

\def\ee           {{\rm e}}

\newcommand{\beq}{\begin{equation}}
\newcommand{\eeq}{\end{equation}}
\newcommand{\beqa}{\begin{eqnarray}}
\newcommand{\eeqa}{\end{eqnarray}}
\newcommand{\beqar}{\begin{eqnarray*}}
\newcommand{\eeqar}{\end{eqnarray*}}

\renewcommand{\eqref}[1]{(\ref{#1})}

\catcode`\@=12

\title{Hydrodynamics of a 5D Einstein-dilaton black hole solution and the corresponding BPS state}
\author[a]{Song He,}
\author[b]{Ya-Peng Hu,}
\author[b]{Jian-Hui Zhang}

\affiliation[a]{Key Laboratory of Frontiers in Theoretical Physics,
Institute of Theoretical Physics, Chinese Academy of Sciences,
Beijing, PRC } \affiliation[b]{Center for High-Energy Physics,
Peking University, Beijing 100871, China}

\emailAdd{hesong@itp.ac.cn}\emailAdd{huyp@pku.edu.cn}\emailAdd{zhangjianhui@gmail.com}

\date{\today}

\abstract{We apply the potential reconstruction approach to generate
a series of asymptotically AdS (aAdS) black hole solutions, with a
self-interacting bulk scalar field. Based on the method, we
reproduce the pure AdS solution as a consistency check and we also
generate a simple analytic 5D black hole solution. We then study
various aspects of this solution, such as temperature, entropy
density and conserved charges. Furthermore, we study the
hydrodynamics of this black hole solution in the framework of
fluid/gravity duality, e.g. the ratio of the shear viscosity to the
entropy density. In a degenerate case of the 5D black hole solution,
we find that the c function decreases monotonically from UV to IR as
expected. Finally, we investigate the stability of the degenerate
solution by studying the bosonic functional energy of the gravity
and the Witten-Nester energy $E_{WN}$. We confirm that the
degenerate solution is a BPS domain wall solution. The corresponding
superpotential and the solution of the killing spinor equation are
found explicitly.}

\begin{document}

\maketitle

\section{Introduction}
  Five-dimensional black hole solutions of Einstein gravity coupled
to dilaton field have been an avenue of intense research for many
years. The gauge/gravity duality
\cite{Maldacena:1997re}\cite{Gubser:1998bc}\cite{Witten:1998qj}\cite{Aharony:1999ti}
promotes finding new kinds of solutions that are asymptotically
$AdS_5$ ($aAdS_5$) in the UV region, which are very important in
building models that can be used to describe the dual field theory.
As we know, the vacuum solution with a negative cosmological
constant is pure $AdS$, which is dual to $\mathscr{N}=4$ super
Yang-Mills (SYM) theory. It is valuable to find new solutions with
asymptotical AdS behavior that may be dual to non-conformal field
theory. Such new solutions will help extend the application of
gauge/gravity duality. Asymptotically AdS solutions can be found for
example in
Refs.\cite{Martinez:2004nb}\cite{Dotti:2007cp}\cite{Hatsuda:2002hz}\cite{Zeng:2009fp}\cite{Torii:2001pg}.
To describe phenomena in realistic field theories, one needs to
break the conformal symmetry. A simple way to do this is to
introduce a nontrivial dilaton. Motivated by this, we introduce a
general framework to generate a series of new $aAdS_5$ solutions
from a bottom-up point of view.

From the phenomenological perspective, introducing the scalar field
and its self interaction to build up effective gravity background is
fairly important for investigating phenomena in field theories via
holography. For example, in holographic QCD (hQCD) communities,
nontrivial dilaton potentials have been used to build dual hQCD
models~\cite{Megias:2010ku,Gubser-T,Gursoy-T}. The authors of
\cite{Gubser-T} \cite{Gursoy-T} make contribution to building up 5D
gravity background in graviton with minimally coupled dilaton system
to describe the QCD phenomena. All the solutions they found are
numerical. Their strategy is to fix the dilaton potential and then
find the gravity solutions. Staring from the 10D string theory or
supergravity, one can use various ways to reduce the theory to the
5D low energy effective gravity theory. After a consistent
truncation or reduction, there could be various versions of black
holes with the scalar fields, such as one charge $\mathcal {N}=4$
black hole
\cite{Behrndt:1998jd}\cite{Cai:1998ji}\cite{Gubser:1998jb},
$\mathcal {N}=2^*$ black hole\cite{Benincasa:2005iv}, 5D effective
quenched background\cite{Benincasa:2006ei} of the Sakai-Sugimoto
model \cite{Sakai:2004cn}. It is a non-trivial task to find gravity
solutions of these theories \cite{ChiouLahanas:2009cs}. The
procedure of solving the gravity background with fixing dilaton
potential as in Refs~\cite{Gubser-T} \cite{Gursoy-T} is difficult,
both analytically and numerically. Is
it possible to reconstruct the whole background in an easy manner? 

Different from the logic of Refs~\cite{Gubser-T} \cite{Gursoy-T}, a bottom-up approach known as
the potential reconstruction approach~\cite{Farakos:2009fx}\cite{Li:2011hp} is indeed a much easier way to obtain
gravity solutions if the information of the metric is known well.
Using this bottom-up approach, a new Schwarzschild-AdS black hole in
five-dimensions coupled to a scalar field was discussed in
\cite{Farakos:2009fx}, while dilatonic black hole solutions with a
Gauss-Bonnet term in various dimensions were discussed in
\cite{Ohta:2009pe}. The new four dimensional gravity solution has
been found in \cite{Kolyvaris:2009pc}.

We will use the potential reconstruction approach to consistently work out the
general 5D gravity solution, which is valuable for studying the phase
transition in thermal QCD phase diagram. In \cite{Li:2011hp}, the authors use this method to
construct a semianalytical gravity solution to study the thermodynamical quantities, their results agree with the numerical results from recent studies in lattice QCD. The authors also study loop operators in
field theory from gauge/gravity duality and the results are
consistent with the lattice QCD calculation. Ref.~\cite{Li:2011hp} provides an excellent example of reconstructing a holographic model using the potential reconstruction approach. It is worthwhile pushing and extending
this approach to reconstruct gravity solutions.

In this paper, we extend the potential reconstruction approach and
study various aspects of a 5D black hole solution generated by this
approach. Motivated from understanding the aAdS solution from
gauge/gravity duality perspective, we have investigated the related
properties of the solution. We first study the thermodynamic quantities
of the solution. In studying the conserved charges associated with
this solution we find several different subtractions rendering
the action finite, which correspond to introducing different
counterterms~\cite{Counterterm,Brown} to the action. We show how the appropriate subtraction can be singled out by looking at a special limit of the solution. In order to
understand the dual hydrodynamical properties containing dilaton
field in the bulk, we also study the ratio of the shear viscosity to
the entropy density from fluid/gravity duality
perspective~\cite{Bhattacharyya:2008jc}. Our result is consistent
with the lowest bound given by using the Green-Kubo formula from
$AdS/CFT$~\cite{Policastro:2001yc,Kovtun:2003wp,Buchel:2003tz,Kovtun:2004de,Mas:2006dy,Son:2006em,Saremi:2006ep,Maeda:2006by,Cai:2008in,Iqbal:2008by,Cai:2008ph,Sinha}.
To investigate the field theory dual to the background, we then
focus on a degenerate case of the 5D black hole solution with
$V_1=0$ and calculate the c function\cite{Alvarez:1998wr}
characterizing the holographic RG flow
\cite{Distler:1998gb}\cite{Freedman:1999gp} of the conformal
anomaly. Finally, we discuss the stability of the degenerate
solution from the bosonic functional energy \cite{Skenderis:1999mm}
and the Witten-Nester energy $E_{WN}$\cite{Witten:1981mf}. The
superpotential for the degenerate solution has been figured out,
which means that the solution is a BPS domain wall solution. From
the Witten-Nester energy $E_{WN}$ point of view, it is also a stable
one. Furthermore, the solution of the killing spinor equation/Witten
equation has been given explicitly. From the results given in this
paper, the new method of reconstructing gravity promotes us to do
further studies in this direction.

 This paper is organized as follows. In Section 2 we briefly
describe the effective approach to reconstruct the gravity
background pointed out in Ref.~\cite{Li:2011hp}. All these gravity
backgrounds satisfy the asymptotical AdS boundary condition near the
UV region. We will point out two strategies to realize the
reconstruction and comment on the approach. In Sections 3 and 4 we
calculate, for a simple 5D black hole solution, the relevant
thermodynamic quantities such as the temperature, the entropy and
the conserved charges, where we also introduce the counterterm
rendering the black hole energy finite in Section 4. In Section 5,
we use the fluid/gravity duality to investigate the ratio of the
shear viscosity to the black hole entropy density. In section 6,
motivated from studies on the holographic RG flow of the conformal
anomaly in an aAdS solution, we calculate the c function for the
degenerate case of the 5D black hole solution and find that it
decreases monotonically from UV to IR. In Section 7, in order to
test the stability of the degenerate aAdS solution, we use the
standard way to investigate the superpotential and work out the
solution of the killing spinor equation. Section 8 is devoted for
conclusion and discussions.

\section{Generating black hole solution for Einstein-dilaton system}
\label{gravitysetup} We start from the minimal non-critical 5D
effective gravity action in the string frame \cite{Li:2011hp}:
\begin{equation} \label{minimal-String-action}
S_{5D}=\frac{1}{16 \pi G_5}\int d^5 x \sqrt{-g^S} e^{-2 \phi}
 \left(R^S + 4\partial_\mu \phi
\partial^\mu \phi- V_S(\phi)\right),
\end{equation}
where $G_5$ is the 5D Newton constant, $g^S$ and $V_S(\phi)$ are the
determinant of the 5D metric and the dilaton potential in the string
frame, respectively. We will list relations between geometrical
quantities in the string frame and in the Einstein frame. One can
easily link the two frames by using the following well-known
relations.

If the metric in the Einstein frame $g^E_{\mu\nu}$ and in the string
frame $g^S_{\mu\nu}$ are connected by the scaling transformation
\begin{equation}
g^S_{\mu\nu} = e^{-2\Omega}g^E_{\mu\nu},
\end{equation}
then the scalar curvature in the Einstein frame and in the string frame has
the following exact relation
\begin{eqnarray}
e^{-2 \Omega} R^S &=& R^E -(D-1)(D-2) \partial_\mu \Omega
\partial^\mu \Omega+ 2(D-1)\nabla^2 \Omega
\end{eqnarray}
with $D$ dimension, and $\nabla^2$ is defined by
$\frac{1}{\sqrt{-g^E}}
\partial_\mu \sqrt{-g^E}
\partial^\mu$, which is useful to derive the exact relation
between the actions in the string frame and in the Einstein frame.

For the special case $D=5$, we have
\begin{eqnarray}
& & \int \sqrt{-g^S} e^{m \Omega} R^S = \int \sqrt{-g^E}
      e^{(m-3)\Omega}
      \left[ R^E -12 \partial_\mu \Omega
      \partial^\mu\Omega+8\nabla^2 \Omega
      \right], \\
& & \int \sqrt{-g^S} e^{m\Omega}\left( g_S^{\mu\nu} \partial_\mu
\Omega
\partial_\nu \Omega\right) = \int \sqrt{-g^E} e^{(m-5)\Omega}\left(
e^{2\Omega}g_E^{\mu\nu} \partial_\mu \Omega \partial_\nu
\Omega\right).
\end{eqnarray}
The dilaton potentials in two different frames are related by
\begin{eqnarray}
V_S(\Omega)=V_E(\Omega) e^{2\Omega}.
\end{eqnarray}
By setting $m=3$ and $\Omega=-\frac{2}{3}\phi$, we have
\begin{equation}
V_S=V_E e^{\frac{-4\phi}{3}}
\end{equation}
and the following relation
\begin{eqnarray}\label{string-enssteinframe}
& & \int \sqrt{-g^S} e^{-2 \phi} \left(R^S + 4\partial_\mu \phi
\partial^\mu \phi- V_S(\phi)\right) \nonumber \\
&=&\int \sqrt{-g^E} \left[ R^E -\frac{4}{3}
\partial_\mu \phi \partial^\mu \phi - V_E(\phi)\right].
\end{eqnarray}
Therefore, the action Eq.(\ref{minimal-String-action}) becomes in the Einstein frame
\begin{equation} \label{minimal-Einstein-action}
S_{5D}=\frac{1}{16 \pi G_5} \int d^5 x
\sqrt{-g^E}\left(R^E-\frac{4}{3}\partial_{\mu}\phi\partial^{\mu}\phi-V_E(\phi)\right).
\end{equation}

We can write down the general ansatz in the string frame which is
similar to holographic QCD models given by \cite{Gursoy-T}
\begin{equation} \label{metric-stringframe}
ds_S^2=\frac{L^2
e^{2A_s}}{z^2}\left(-f(z)dt^2+\frac{dz^2}{f(z)}+dx^{i}dx^{i}\right)
\end{equation}
with $L$ the radius of ${\rm AdS}_5$. The metric in the string frame
is useful to calculate the loop operator in holographic QCD
communities. To derive the Einstein equations and to study the
thermodynamical quantities, we transform it to the Einstein
frame
\begin{eqnarray} \label{metric-Einsteinframe}
ds_E^2=\frac{L^2 e^{2A_s-\frac{4\phi}{3}}}{z^2}\left(-f(z)dt^2
+\frac{dz^2}{f(z)}+dx^{i}dx^{i}\right).
\end{eqnarray}

The general Einstein equations and equation of motion of scalar from
the action (\ref{minimal-Einstein-action}) take the form of
\begin{eqnarray} \label{EOM}
E_{\mu\nu}+\frac{1}{2}g^E_{\mu\nu}\left(\frac{4}{3}
\partial_{\mu}\phi\partial^{\mu}\phi+V_E(\phi)\right)
-\frac{4}{3}\partial_{\mu}\phi\partial_{\nu}\phi
=0.\nonumber\\\label{EOMS} \Box \phi =\frac{3}{8}\frac{\partial
V_{E}}{\partial \phi }.
\end{eqnarray} Where $\Box$ is the d'Alembertian operator in curved space.
By using Eqs (\ref{minimal-Einstein-action}) and
(\ref{metric-Einsteinframe}), we can derive the following nontrivial
Einstein equations in $(t,t), (z,z)$ and $(x_1, x_1)$ spaces,
respectively\bea\label{equation-graviton1} &{}&b''(z)+\frac{b'(z)
f'(z)}{2 f(z)}+\frac{2}{9} b(z) \phi '(z)^2+\frac{b(z)^3 V_E(\phi
(z))}{6 f(z)}=0,\\\label{equation-graviton2}&{}& \phi '(z)^2-\frac{9
b'(z)^2}{b(z)^2}-\frac{9 b'(z) f'(z)}{4 b(z) f(z)}-\frac{3 b(z)^2
V_E(\phi (z))}{4
f(z)}=0,\\\label{equation-graviton3}&{}&f''(z)+\frac{6 b'(z)
f'(z)}{b(z)}+\frac{6 f(z) b''(z)}{b(z)}+\frac{4}{3} f(z) \phi
'(z)^2+b(z)^2 V_E(\phi (z))=0,\eea where $b(z)=\frac{L
e^{A_E}}{z} $ and $A_E(z)=A_s(z)-\frac{2}{3}\phi(z)$.

One should notice that the above three equations are not
independent. There are only two independent functions, one can solve them from two of the three equations, and the third equation can be used to check the consistency of the solutions. To simplify
the procedure of finding solution, one can use the following two
equations without dilaton potential $V_E$
\begin{eqnarray}\label{AF}
 &{}&A_s''(z)+A_s'(z) \left(\frac{4 \phi '(z)}{3}+\frac{2}{z}\right)-A_s'(z)^2-\frac{2 \phi ''(z)}{3}-\frac{4 \phi '(z)}{3
 z}=0,\\\label{ff}
 &{}&f''(z)+ f'(z)\left(3 A_s'(z)-2 \phi '(z)-\frac{3 }{z}\right)=0.
\end{eqnarray}
These two equations can be obtained from (\ref{equation-graviton1})(\ref{equation-graviton2})(\ref{equation-graviton3})
by canceling terms involving $V_E$. Eq.(\ref{AF}) is our starting point
to find gravity solution. The EOM of the dilaton field is given
explicitly as following
\begin{equation}
\label{fundilaton} \frac{8}{3} \partial_z
\left(\frac{L^3e^{3A_s(z)-2\phi} f(z)}{z^3}
\partial_z \phi\right)-
\frac{L^5e^{5A_s(z)-\frac{10}{3}\phi}}{z^5}\partial_\phi V_E=0.
\end{equation}
Our first strategy of reconstructing the gravity background is using as an input the conformal factor $A_s(z)$ and
determine the function $\phi(z)$. Having $A_s$ and $\phi$,
we can solve for the unknown functions $f(z)$ and $V_E(z)$ from Eqs (\ref{AF}) and (\ref{fundilaton}). The second strategy is to choose a special form for $\phi$ and then determine the conformal
factor $A_s$ from Eq.~(\ref{AF}). The remaining steps is the same as in the first strategy. Therefore,
one can use the two kinds of generating function $A_s$, $\phi$ to
reconstruct the gravity solution as we expect. Here we list the
general formalism of generating gravity solution by using $A_s$ as an input.
For the second strategy, we do not find general a formalism to
generate solutions here. We should comment that it is also possible
to find analytical or numerical solution by using the second
strategy.

Starting with a given geometric structure $A_s(z)$, we can derive the
general solutions to the Einstein equations, which take the
following form
\begin{eqnarray} \label{general-solution1}
 \phi(z)&=&\phi _ 0 + \phi _ 1\int_ 0^
 z\frac {e^{2 A_s(x)}} {x^2}\, dx + \frac {3 A_s(z)} {2} \nonumber \\
  &+ & \frac{3}{2}\int_ 0^
    z\frac {e^{2 A_s(x)}\int _ 0^{x}y^2 e^{-2 A_s(y)} A_s'(y) {}^2 dy} {x^2}\, dx,\\\label{general-solution2}
f(z)&=&f_0+f_ 1\left (\int_ 0^z x^3 e^{2\phi (x) - 3 A_s (x)}\,
dx\right),\\ \label{general-solution3} V_E(\phi)&=&\frac {e^{\frac
{4\phi (z)}{3} - 2
A_s (z)}}{L^2} \nonumber \\
   & & \left (z^2 f^{''}(z) - 4 f (z)\left (3 z^2 A_s''(z) - 2
z^2\phi^{''}(z) + z^2\phi' (z)^2 + 3 \right) \right),
\end{eqnarray}
where $\phi_0,\phi_1,f_0,f_1$ are constants of integration. As a
consistency check, one can set $A_s=0$, $\phi$ is then found to be $\phi=\phi_0$ with $\phi_0$ a constant, $\phi_1=0$ is indicated by requiring a regular solution for $\phi$. In terms of
the (\ref{general-solution2}), one can see the general system can
reproduce the pure $AdS_5$ vacuum solution if $f_0=1$ and $f_1=0$.
Plugging these values to Eq.~(\ref{general-solution3}),
one can easily get the scalar potential $V_s(z)={12\over L^2}$, as one
expects for the pure $AdS_5$ case.

From the results above, one can easily obtain various gravity
solutions by putting an $A_s(z)$ into
Eqs~(\ref{general-solution1})(\ref{general-solution2})(\ref{general-solution3})
and imposing physical boundary conditions near UV and near IR
region. Here we should point out that, in contrast to the ordinary
procedure of finding the gravity solution after fixing the bulk
potential $V_s$ or $V_E$, the framework motivated from the bottom-up
point of view is good at producing lots of, and also helpful to
reconstruct the gravity background. That is to say that one is able
to extract suitable backgrounds to describe the phenomena on the
field theory side. It is an efficient way of linking gravity and
field theory. This approach will shed light on the application of
gauge/gravity duality from a bottom-up viewpoint.

Before ending this section, we will comment on this approach. One
should note from the solution in Eqs (\ref{general-solution1})
(\ref{general-solution2}) (\ref{general-solution3}) that the
geometric parameters such as $\phi_0,\phi_1, f_0,f_1$ contribute to
the dilaton potential $V_E(\phi_0,\phi_1, f_0,f_1)$. Changing these
parameters in the potential $V_E$ means that the theory is changed.
In other words, different values of the parameters $\phi_0,\phi_1,
f_0,f_1$ in $V_E$ correspond to different gravity theories. This is
a main difference between this method and the ordinary procedure of
finding the gravity solution with a fixed potential $V_E$. In some
gravity solutions \cite{Li:2011hp}\cite{He:2010ye} reconstructed by
this method, it seems possible that these different theories can be
linked by the same form of action with different values of
parameters in $V_E$.  The different values of parameters correspond
to different configurations of bulk field and different potentials,
therefore these theories are not equivalent to each other.  For one
simple solution given in Appendix A of Ref.~\cite{Li:2011hp},
$V_1\neq 0$ in $f(z)$ can affect the form of potential through
(\ref{general-solution3}). For the dilaton potential with only a
cosmological constant, there are two gravity solutions, the pure
$AdS_5$ solution and the $AdS$ Schwarzschild black hole. This means
different parameter $f_1$ is dual to different states of one gravity
theory in this case. Once one constructs the gravity background, one
should confirm that the potential $V_E$ and $A_s(z), f(z)$ satisfy
the constraints from other physical considerations, e.g. the
Breitenlohner-Freedman (BF) bound of scalar field near AdS boundary,
the finiteness of the action, a well-defined boundary of the system
and so forth. In general, using this effective approach necessitates
some physical checks to ensure that the solution is self-consistent.

\section{Black hole solution and thermodynamic quantities}
In this section, we just introduce a simple analytical gravity
solution and then focus on the properties of
the solution. If one is interested in other solutions, please refer to the appendix A of Ref.~\cite{Li:2011hp}. Before giving the 5D black hole solution
generated by the above method from Eqs
(\ref{general-solution1})(\ref{general-solution2})(\ref{general-solution3}),
we present the relation of the geometric factors $A_s$ and
$A_E$ in the string frame and in the Einstein frame
\begin{eqnarray} \label{Ametric-Einsteinframe}
ds_E^2&=&\frac{L^2 e^{2A_s-\frac{4\phi}{3}}}{z^2}\left(-f(z)dt^2
+\frac{dz^2}{f(z)}+dx^{i}dx^{i}\right)\nonumber\\
&=&\frac{L^2 e^{2A_E}}{z^2}\left(-f(z)dt^2
+\frac{dz^2}{f(z)}+dx^{i}dx^{i}\right)
\end{eqnarray}
with $A_E(z)=A_s(z)-\frac{2\phi(z)}{3}$.

We now give a simple analytical 5D black hole solution which is first obtained in Ref.~\cite{Li:2011hp}
\begin{eqnarray}\label{solutionexact1}
A_E(z)&=&\log \left(\frac{z }{z_0\sinh(\frac{z}{z_0})}\right),\nonumber\\
\phi(z)&=&\frac{3 z}{2 z_0},\nonumber\\
f(z)&=&1-\frac{V_{1}}{3}(2-3\cosh(\frac{z}{z_0})+\cosh^3(\frac{z}{z_0})).
\end{eqnarray}
 The non-trivial dilaton potential in Eq.~(\ref{string-enssteinframe}) is given by \bea\label{dilatonpotential1}
V_E(\phi)&=&-\frac{16V_{1}\sinh^6(\frac{\phi}{3})+9\sinh^2(\frac{2\phi}{3})+12}{L^2},
\eea where $z_0$ is an integration constant and $V_{1}$ is the
constant parametrizing the dilaton potential. The special solution
with $V_1=0$ thus $f(z)=1$ is a degenerate aAdS solution without
black hole, we will discuss it later and show that it corresponds to
a BPS domain wall solution in the context of supergravity.

We now discuss the relevant thermodynamic quantities for the black
hole solution (\ref{solutionexact1}). The first thing is how to
parameterize the Hawking temperature, which is defined as
$\left.\frac{f'(z)}{4\pi}\right|_{z_h}$. A black-hole solution with
a regular horizon is characterized by the existence of a surface
$z=z_h$ satisfying $f(z_h)=0$. The Euclidean version of the solution
is defined only for $0 < z < z_h$, in order to avoid the conical
singularity, the periodicity of the Euclidean time can be fixed by
\begin{equation}
\tau \rightarrow \tau+\frac{4\pi}{|f'(z_h)|}.
\end{equation}
This determines the temperature of the solution as \bea\label{temp}
T &=&\frac{|f'(z_h)|}{4\pi}\nonumber\\
&=&{1\over{ 4\pi}}\left| V_1 \sinh {z_h\over z_0}- V_1 \cosh^2 {z_h\over
z_0}\sinh {z_h\over z_0}\right|, \eea where $V_1$ can be expressed by
the horizon of the black hole $z_h$ as
 \bea V_1=
\frac{3}{2-3\cosh(\frac{z_h}{z_0})+\cosh^3(\frac{z_h}{z_0})}.\eea

Following the standard Bekenstein-Hawking formula \cite{entropy-BK},
one can easily read the black-hole entropy density $s$ from the geometry given in Eq.(\ref{metric-Einsteinframe}), which is defined by
the area $A_{area}$ of the horizon as
\begin{equation}
\label{entrpy} s={\frac{A_{area}}{4 G_5 V_3}=
\frac{L^3}{4G_5}\left(\frac{e^{A_s-\frac{2}{3}\phi}}{z}\right)^3}\Bigg|_{z_h},
\end{equation}
where $G_5 $ is the Newton constant in 5D curved space and $V_3$ is
the volume of the spatial directions. It is noticed that the entropy
density is closely related to the metric in the Einstein frame.

For later convenience, we set $L=1$ and make the coordinate
transformations $z=1/r, z_0=1/r_0$. The
metric~(\ref{Ametric-Einsteinframe}) then becomes
\begin{equation}
ds^{2}=r_0^2\mbox{csch}^2\left(\frac{r_0}{r}\right)\big(-f(r)dt^2+\frac{1}{%
r^{4}}\frac{dr^{2}}{f(r)}+dx_{i}^{2}\big), \label{solutiontr}
\end{equation}
Setting $R=r_0\,\mbox{csch}\left({r_0\over r}\right)$, one has
\begin{equation}
ds^{2}=R^2\Big(-f(R)dt^2+\frac{1}{%
R^{2}(R^2+r_0^2)}\frac{dR^{2}}{f(R)}+dx_{i}^{2}\Big).
\label{solutiontR}
\end{equation}

\section{The conserved charges and counterterms}
\label{section4} In this section, we will calculate the conserved
charges of the solution (\ref{solutiontR}) by using the counterterm
method~\cite{Counterterm}. This method is based on the Brown-York
quasilocal stress tensor $T_{ab}$ which is first proposed
in~\cite{Brown}. According to this method, for a $d+1$ dimensional
spacetime $\mathcal{M}$ with the boundary geometry~$\partial
\mathcal {M}$, the metric $\gamma _{a b }$ on the boundary $\partial
\mathcal {M}$ can be written with the help of an ADM decomposition as
\begin{equation}
\gamma _{a b }dx^{a }dx^{b }=-N_{\Sigma }^{2}dt^{2}+\sigma
_{ij}(dx^{i}+N_{\Sigma }^{i}dt)(dx^{j}+N_{\Sigma }^{j}dt),
\label{BoundaryADM}
\end{equation}
where $\sigma_{ij}$ is the induced metric on the spacelike
hypersurface $\Sigma $ of the boundary $\partial \mathcal {M}$, and
$N_{\Sigma }$ is the lapse function related to the timelike normal
vector $u^{a }$ of $\Sigma $ as $u^{a }=(\frac{\partial}{\partial
t})^a/N_{\Sigma }$, $N_{\Sigma }^{i}$ is the shift vector field.
Therefore, the conserved charge $Q_{\xi }$ associated with the
killing vector $\xi ^{a }$ is defined by the quasilocal stress
tensor $T_{ab}$ as~\cite{Counterterm,Brown,Hu:2009rj}
\begin{equation}
Q_{\xi }=\int_{\Sigma }d^{d-1}x\sqrt{\sigma }(u^{a }T_{a b }\xi ^{b
}),   \label{Conservedcharge}
\end{equation}%
where $\sigma$ is the determinant of $\sigma _{i j }$, the
energy associated with the timelike killing vector $\xi ^{a }$ and the
momentum could be defined respectively as
\begin{eqnarray}
E &=&\int_{\Sigma }d^{d-1}x\sqrt{\sigma }N_{\Sigma }(u^{a }T_{a b
}u^{b }), \label{Mass}
\\P_{i} &=&\int_{\Sigma }d^{d-1}x\sqrt{\sigma }\sigma _{ij}u_{a
}T^{j a }.  \label{Momentum}
\end{eqnarray}
We have used $a,b$ to denote indices on the boundary $\partial \mathcal {M}$, to avoid confusion from $\mu,\nu$ used in previous sections denoting indices in the $d+1$ bulk
spacetime $\mathcal{M}$. $i,j$ are indices on the spacelike hypersurface $\Sigma $.

For the asymptotically $AdS_{5}$ solution, the quasilocal stress
tensor usually is~\cite{Counterterm}
\begin{equation}
T_{a b }=\frac{1}{8\pi }(\theta _{a b }-\theta \gamma _{a b }-%
3 \gamma _{a b }-G_{a b }),  \label{stresstensor}
\end{equation}%
where $\theta _{a b }=-\frac{1}{2}(\nabla _{a }n_{b }+\nabla
_{b}n_{a })$ is the extrinsic curvature of the boundary $\partial
\mathcal {M}$, $n^{a }$ is the normal vector of $\partial \mathcal
{M}$, and $G_{a b }$ is the Einstein tensor of $\gamma _{a b}$, and
the last two terms in (\ref{stresstensor}) come from the counterterm
action~\cite{Counterterm}. However, in our case in
(\ref{solutiontR}), we find that the quasilocal stress
tensor (\ref{stresstensor}) defined as above would be divergent, which is common
in Einstein-dilaton gravity. There have been many references
\cite{Nojiri:1998dh,Cai:1999xg,Nojiri:2000kh} attacking the issue of removing the divergence and achieving a well-defined quasilocal stress tensor.
In our case (\ref{solutiontR}), we found two renormalizations that remove the divergence.
In the first scheme we have the following well-defined quasilocal stress tensor
\begin{equation}
T_{a b }=\frac{1}{8\pi }(\theta _{a b }-\theta \gamma _{a b }-(1-\frac{V_E(\phi)}{6})%
\gamma _{a b }-G_{a b }),  \label{stresstensor1}
\end{equation}%
while in the second
\begin{equation}
T_{a b }=\frac{1}{8\pi }(\theta _{a b }-\theta \gamma _{a b }-(-1+\sqrt{-\frac{4V_E(\phi )}{3}})%
\gamma _{a b }-G_{a b }).  \label{stresstensor2}
\end{equation}%
If one expands the contribution of two different counterterms in Eqs
(\ref{stresstensor1}) and (\ref{stresstensor2}), one finds
\bea\label{scheme1} 1-\frac{V_E(\phi)}{6}&&\sim
3+\frac{3r_0^2}{2R^{2}}+\mathcal
O(\frac{1}{R^5}),\nonumber\\\label{scheme2}
-1+\sqrt{-\frac{4V_E(\phi )}{3}}&&\sim
3+\frac{3r_0^2}{2R^{2}}-\frac{9r_0^4}{32R^{4}}+\mathcal
O(\frac{1}{R^5}). \eea

Actually we can write a most general well-defined quasilocal stress
tensor as
\begin{equation}
T_{a b }=\frac{1}{8\pi }(\theta _{a b }-\theta \gamma _{a b
}-X(\phi) \gamma _{a b }-G_{a b }), \label{stresstensor3}
\end{equation}%
where $X(\phi)$ is associated to the counterterm,  and behaves as $X(\phi)\rightarrow
3+\frac{3r_0^2}{2R^{2}}+\frac{C_{4}}{R^{4}}+\mathcal
O(\frac{1}{R^5})$ as $R$ goes to infinity ($C_{4}$ is a
constant), in order to achieve a finite $T_{ab}$. The quasilocal stress tensor in
Eq.~(\ref{stresstensor1}) clearly corresponds to the minimal subtraction
with $C_4=0$.
 In terms of the general form of the stress tensor
 Eq.~(\ref{stresstensor3}), the useful components of quasilocal
 stress tensor as $R$ goes to infinity are
\begin{eqnarray}
T_{tt}&=&\frac{(3+3V_{1})r_{0}^4+8C_{4}}{8R^2}, \nonumber\\
T_{xx}&=&T_{yy}=T_{zz}=\frac{(-3+V_{1})r_{0}^4-8C_{4}}{8R^2}.
\label{Stresstensor3Useful}
\end{eqnarray}
Where $V_3$ is the volume of space including $x_1,x_2,x_3$.
Therefore, the energy of black hole solution (\ref{solutiontR}) can
be expressed as
\begin{eqnarray}
E&=&\frac{(3+3V_{1})r_{0}^4+8C_{4}}{8}V_{3}.\label{ConservedCharges}
\end{eqnarray}
At first sight, the $C_4$ term seems to introduce an ambiguity. However, as we will show in the next section,
this ambiguity can actully be removed by looking at some special limit of our solution.

\section{Hydrodynamics of dual boundary fluid via AdS/CFT correspondence}
\label{hydro} In this section we study the hydrodynamics of the
boundary fluid dual to the bulk gravitational solution
(\ref{solutiontR}), following the method proposed by
Ref.~\cite{Bhattacharyya:2008jc}. In
Ref.~\cite{Bhattacharyya:2008jc} the authors established a
systematic way to map the hydrodynamic expansion of the boundary
theory to the gradient expansion of the bulk gravity. One starts
from a static black brane solution boosted along the translational
invariant spatial directions, which corresponds to the boundary
dynamics in global thermal equilibrium. The parameters
characterizing the bulk solution such as the black hole temperature,
velocities correspond precisely to the hydrodynamic degrees of
freedom. To study the first order hydrodynamics, one moves away from
the equilibrium by promoting the parameters to slowly-varying
functions of the boundary coordinates, the original black brane
solution no longer fulfills the equation of motion of the bulk
gravity. An exact solution of the bulk gravity can be achieved only
if the parameters now satisfy a set of equations of motion, which
turn out to be the equations of the boundary fluid dynamics. In this
way one establishes a one-to-one mapping between the bulk
gravitational solution and the boundary fluid dynamics.

To start with, we transform the metric in Eq.~(\ref{solutiontR}) into the
Eddington-Finkelstein coordinates $v=t+r_*$ with
$dr_*=\frac{dR}{R\sqrt{R^2+r_0^2}f(R)}$ \beq
ds^2=-R^2 f(R)dv^2+\frac{2R}{\sqrt{R^2+r_0^2}}dv dR+R^2 dx^i dx_i.
\eeq

Following the spirit of Ref.~\cite{Bhattacharyya:2008jc}, we start
from the boosted form of the above metric
\beq\label{boostedsol} ds^2=-R^2 f(R)(u_a
dx^a)^2-\frac{2R}{\sqrt{R^2+r_0^2}}u_a dx^a dR+R^2
P_{ab}dx^a dx^b \eeq with
\begin{equation}
u^v = \frac{1}{ \sqrt{1 - \beta_i^2} }~~,~~u^i = \frac{\beta_i}{
\sqrt{1 - \beta_i^2} },~~P_{ab}= \eta_{ab} + u_a u_b,
\end{equation}
where $x^a=(v,x_{i})$ denote the coordinates on the boundary, boost velocities $\beta^i $ are constants,
$P_{ab}$ is the projector onto spatial directions, and the
indices in the boundary are raised and lowered with the Minkowski
metric $\eta_{ab}$.

According to the procedure outlined above, we now lift the
parameters in the metric in Eq.~(\ref{boostedsol}) to slowly-varying
functions of $x^a$. The metric no longer
fulfills the equations of motion (\ref{EOM}). To find the solutions, we can define the following quantities
\begin{eqnarray}\label{W1}
W_{\mu \nu }&=&R_{\mu \nu }-\frac{1}{3}g_{\mu \nu }V_{E}(\phi
)-\frac{4}{3}\partial _{\mu }\phi \partial _{\nu }\phi, \\\label{W2}
W&=&\square \phi -\frac{3}{8}\frac{\partial V_{E}}{\partial \phi },
\end{eqnarray}
where vanishing of the RHS of Eqs (\ref{W1}) (\ref{W2}) will give
the equations of motion (\ref{EOM}). After lifting the parameters
$\beta^i, r_0$ to functions of the boundary coordinates $x^a$,
$W_{\mu \nu }$ and $W$ are proportional to the derivatives of
$\beta^i, r_0$.  They give rise to the source terms on the LHS of
(\ref{W1}) (\ref{W2}). We should introduce corrections to the metric
and to the dilaton field, such that these corrections compensate for
the effect of the source terms $W_{\mu\nu}, W$. Then a solution of
the bulk equations of motion can still be achieved. We will consider
the derivative expansion to the first order. As usual, there exists
a gauge redundancy in the metric. To eliminate this redundancy, we
choose the background field gauge for the metric following
Ref.~\cite{Bhattacharyya:2008jc}. Note that the background metric
has a spatial $SO(3)$ symmetry, it allows us to write the
corrections to the metric in terms of the $SO(3)$ scalar, vector and
traceless symmetric tensor components.

To the first order, we have
\begin{align}\label{correction}
{ds^{(1)}}^2 &= \frac{ k(R)}{R^2}dv^2 + \frac{2R\,h(R)}{\sqrt{R^2+r_0^2}}dv dR + 2 \frac{j_i(R)}{R^2}dv dx^i +R^2 \Big(\alpha_{ij}(R) -\frac{2}{3} h(R)\delta_{ij}\Big)dx^i dx^j, \nonumber\\
\phi^{(1)} &= a(R).
\end{align}
where $h(R), k(R)$ are the scalar components, $j_i$ and
$\alpha_{ij}(R)$ the vector and tensor components, respectively.
$a(R)$ corresponds to correction term of dilaton field.

The equations of motion for the correction terms can be derived by
requiring that the bulk equations of motion are fulfilled after
adding the correction terms to the metric. We are interested in
extracting the shear viscosity of the dual boundary fluid. For this
purpose, we only need the solution for the non-diagonal components
of the traceless symmetric tensor $\alpha_{ij}$. The equation for
such components $\alpha_{ij} (i\ne j)$ can be obtained
 by putting (\ref{correction}) into (\ref{W1})(\ref{W2}), which reads
\beq\label{EOMalphaij}
c_2(R)\alpha_{ij}''(R)+c_1(R)\alpha_{ij}'(R)-9(\partial_i\beta_j+\partial_j\beta_i)=0
\eeq with
\begin{align}
c_2(R)&=R^2\sqrt{R^2+r_0^2}(-3+2V_1)-\frac{(2R^4+R^2 r_0^2-r_0^4)V_1}{R},\nonumber\\
c_1(R)&=-10R^2 V_1-3r_0^2 V_1+\frac{r_0^4
V_1}{R^2}+\frac{R(5R^2+4r_0^2)(-3+2V_1)}{\sqrt{R^2+r_0^2}}.
\end{align}
Note that $\alpha_{ij}(R)$ has the following form \beq
\alpha_{ij}(R)=\alpha(R)(\partial_i\beta_j+\partial_j\beta_i-\frac{2}{3}\delta_{ij}\partial_k\beta^k),
\eeq where the tensor structure on the RHS is the traceless
symmetric tensor structure constructed from the derivative of
$\beta_i$.

Solving Eq.~(\ref{EOMalphaij}), we find the following asymptotic
form for $\alpha(R)$ \beq\label{alphaasympt} \alpha(R)\sim
\frac{1}{R}-\frac{r_0^2}{6R^3}-\frac{r_H^3}{4 R^4}+\mathcal
O(\frac{1}{R^5}). \eeq The equations for the remaining correction
terms in Eq.~(\ref{correction}) are rather lengthy, we do not list
those equations here, but we give the asymptotic form of those
remaining correction terms for completeness
\begin{align}
a(R)&\sim \big(\frac{3}{2R^2}-\frac{3r_0^2}{R^4}\big)(\partial_v r_0+\frac{r_0\partial_i\beta^i}{3})+\mathcal O(\frac{1}{R^6}), \nonumber\\
h(R)&\sim \big(-\frac{4r_0}{R^3}-\frac{42 r_0^3}{R^5}\big)(\partial_v r_0+\frac{r_0\partial_i\beta^i}{3})+\mathcal O(\frac{1}{R^7}), \nonumber\\
k(R)&\sim \frac{2R^3\partial_i\beta^i}{3}-R r_0(r_0\partial_i\beta^i+2\partial_v r_0)-\frac{r_0^3(109r_0\partial_i\beta^i+372\partial_v r_0)}{60R}+\mathcal O(\frac{1}{R^3}), \nonumber\\
j_i(R)&\sim R^3\partial_v\beta_i+\frac{R r_0(2\partial_i
r_0-r_0\partial_v\beta_i)}{6}+\mathcal O(\frac{1}{R}),
\end{align}
where in obtaining these results and the asymptotic form of
$\alpha_{ij}$ we have used the same boundary conditions and gauge choice as in
Refs.~\cite{Bhattacharyya:2008jc,Hur:2008tq,Hu:2011ze}, e.g. non-deformation of the field theory metric and working in the Landau frame.

Having the results for the correction terms, we can derive the boundary stress tensor from the corrected metric. This can be done by varying the total
action with respect to the boundary metric $\gamma_{ab}$.
From~(\ref{stresstensor3}), the boundary stress tensor is given by
\beq\label{Tmunu}
 T_{ab}=\frac{1}{8\pi }(\theta _{ab}-\theta \gamma _{ab}-X(\phi)
\gamma _{ab}-G_{ab}). \eeq

Note that the background metric upon which the dual field theory
resides is $h_{ab}=\lim_{R \rightarrow \infty}
\frac{1}{R^2}\gamma_{ab}$, which is the Minkowski metric, the
expectation value of the stress tensor of the dual fluid $\tau
_{ab}$ can then be computed from~\cite{Myers:1999psa}
\begin{eqnarray}\label{relation}
\label{Tik-CFT} \sqrt{-h}h^{ab}<\tau _{bc}>=\lim_{R\rightarrow
\infty }\sqrt{-\gamma }\gamma ^{ab}T_{bc},
\end{eqnarray}
where $T_{ab}$ is the boundary stress tensor in Eq.~(\ref{Tmunu}).
On the other hand, the stress tensor of the boundary fluid has the
following form \beq \tau_{ab}=P(\eta_{ab} +u_a u_b)+\rho\, u_a u_b -
2 \eta \sigma_{ab}, \label{StressTensor} \eeq
where
\begin{align}
\sigma ^{ab}&\equiv \frac{1}{2} P^{ac} P^{bd } \left(\partial _{c
}u_{d }+\partial _{d }u_{c }\right)-\frac{1}{3} P^{ab} \partial _{c
}u^{c }, \nonumber\\
P&=\frac{(-3+V_{1})r_0^4-8C_{4}}{8},\nonumber\\
\rho&=\frac{(3+3V_{1})r_0^4+8C_{4}}{8}.
\end{align}

From the above equation it is clear that, to read off the shear
viscosity $\eta$, we need only the non-diagonal components of
$T_{ab}$, which give \beq
T_{ij}=-\frac{r_H^3(\partial_i\beta_j+\partial_j\beta_i)}{16\pi
R^2}. \label{Tij} \eeq Thus from (\ref{relation}),
(\ref{StressTensor}) and (\ref{Tij}), the shear viscosity can be
read off $\eta=r_H^3/16\pi $, and the ratio of the shear viscosity
and the entropy density is
\begin{equation}
\label{ror}
 {\eta \over s} = {r_H^3/(16\pi)\over r_H^3/4} = {1 \over 4 \pi},
\end{equation}
which agrees with the result obtained previously from other
considerations\cite{Iqbal:2008by,Cai:2008ph}.

A few comments on the $C_4$ term are in order. First of all, we do
not include the bulk viscosity term in Eq.~(\ref{StressTensor}),
since we have checked that it vanishes in our case for all choices
of $C_{4}$. It should be pointed out that this result is nontrivial,
because usually the bulk viscosity term can appear when there is a
dilaton field in the bulk. Whether the zero bulk viscosity is
related to our special black hole solution (\ref{solutionexact1}) or
to the special dilaton potential (\ref{dilatonpotential1}) is
unclear at this stage, and deserves further studies. Second, the
potential ambiguity related to the choice of $C_4$ can be removed by
investigating a special limit of our solution. Note that when
$r_0=0$, our solution becomes pure $AdS_{5}$. If one agrees that the
energy-momentum of the field theory dual to pure $AdS_{5}$ vanishes,
e.g. as in Ref.~\cite{Counterterm}, $C_4$ has to be zero. This
corresponds to the minimal subtraction. The trace of the
energy-momentum tensor is then given by $\tau=-\frac{3r_0^4}{2}$,
which implies a nonzero conformal anomaly unless $r_0=0$, which is
pure $AdS$.

\section{Degenerate aAdS, domain wall and the c-function}
Our purpose in this section is to show that the degenerate aAdS
solution with $V_1=0$ is a domain wall solution, and to demonstrate
a simple property of the bulk that translates via holography into a
c-theorem \cite{Alvarez:1998wr} for the boundary theory. Based on
the action (\ref{minimal-Einstein-action}), the general equations of
motion \cite{Cvetic:1999fe} for graviton and scalar field are
\bea\label{generalEOM1} R_{\mu\nu}&=&{4\over3}\partial_\mu
\phi\partial_\nu\phi+{1\over {3}}g_{\mu\nu}
V_E(\phi),\\\label{generalEOM2} \Box \phi
&=&\frac{3}{8}\frac{\partial V_{E}}{\partial \phi },\eea  where we
have used the fact $R=\frac{4}{3}(\partial\phi)^2-{5\over
3}V_E(\phi)$ to simplify the formula (\ref{EOM}). In order to
consider the holographic RG flow of conformal anomaly, we will
discuss the c function \cite{Henningson:1998gx}, under the domain
wall ansatz for convenience.
 Let us begin by making the domain wall ansatz \cite{Cvetic:1999fe}
\bea\label{domain-wall} ds^2=
e^{2 B(r)} ds^2(\mathbb{E}^{1,3})+ dr^2 \nonumber\\
=U^2  ds^2(\mathbb{E}^{1,3})+{1\over (\partial_r B)^2}{dU^2\over
U^2}\eea with dilaton field $\phi(r)$. An important aspect of the
AdS/CFT correspondence is the notion that the radial coordinate $U$
of AdS can be regarded as a measure of energy. In order to connect
with our previous ansatz (\ref{metric-Einsteinframe}), we make the coordinate transformation as $dr =-\frac{ L
e^{A_E(z)}}{{z}} dz $ and impose the condition
$B(r)={\left(A_E(z)-\log({z\over L})\right)}$. The new form of the
Einstein-dilaton equations for the metric (\ref{domain-wall}) can be
reduced to the following form \bea
&&12(\partial_r B)^2 - {4\over 3}(\partial_r \phi)^2 +  V_E(\phi)=0, \label{secorder1} \\
&&3 \partial_r^2 B + 6 (\partial_r B)^2
+ {2\over 3}(\partial_r \phi)^2 + V_E(\phi) =0, \label{secorder2} \\
&&\partial_r^2 \phi + 4 \partial_r B \partial_r \phi - {3\over
8}V_E'(\phi) = 0, \label{secorder3} \eea where the prime indicates
the differentiation with respect to $\phi$. One can use the
coordinate transformation to reproduce the equations
(\ref{equation-graviton1})(\ref{equation-graviton2})(\ref{equation-graviton3})
as a consistency check. One should note that the above three
equations are not independent, as in our previous case given in
section \ref{gravitysetup}. The above equations
(\ref{secorder1})(\ref{secorder2})(\ref{secorder3}) imply the
following simple exact relation
  \bea \label{sec4}
\partial_r^2 B &=&  -{4 \over 9} (\partial_r \phi)^2\nonumber\\
&=&-{4\over 9}\left(\frac{\partial z}{\partial r}\partial_z
\phi(z)\right)^2 =-{4\over 9L^2}\left(\frac{z}{e^{A_E(z)}}\partial_z
\phi(z)\right)^2. \eea The $c$ function proposed by the author of
\cite{Henningson:1998gx} is useful for studying the conformal
anomaly in this system. In terms of the definition, the c function
is a positive function of the coupling constants that is
non-increasing along the RG flow from the UV region
($U\rightarrow\infty, z\rightarrow0$) to the IR region
($U\rightarrow 0, z\rightarrow\infty$). The definition of the c
function is
\cite{Alvarez:1998wr}\cite{Distler:1998gb}\cite{Freedman:1999gp}
\bea \label{cfun}\mathscr{C}\equiv \int \sqrt{-\gamma}<T>=
{\mathscr{C}_0\over (
\partial_r B(r))^3},\eea where $T$ is the the trace of energy momentum tensor of dual field thery and $\gamma$ is the determinant of the metric on the boundary. The important thing is that once one has fixed
$\mathscr{C}_0$ within a particular supergravity theory, the formula
(\ref{cfun}) gives a form for computing the anomaly coefficients in
any conformal theory dual to an $AdS$ vacuum of that
supergravity theory. The sign of $\mathscr{C}_0$ ensures that
$\mathscr{C}$ is a positive function. Now we consider a
special spacetime (\ref{solutionexact1}) with $V_1=0$ which is
asymptotic to $AdS$ spaces for $z\rightarrow 0$ or
$r\rightarrow \infty$. Here we can get that \bea U
\frac{\partial}{\partial_U} \mathscr{C} &=& -3
\mathscr{C}\frac{\partial^2 B}{(\partial_r B)^2}\nonumber\\&=&
 \frac{4 \mathscr{C}}{3 L^2}\left(\frac{\partial_z
\phi(z)}{\partial_z\frac{\exp({A_E(z)})}{z}} \right)^2
\nonumber\\&=&
 \frac{4 }{3 L^2} {\mathscr{C}_0\over \coth^3 ({z\over z_0})} \left(\frac{\partial_z
\phi(z)}{\partial_z\frac{\exp({A_E(z)})}{z}} \right)^2 \geq 0 \eea
by using the function (\ref{sec4}). It is easy to see that the
$\mathscr{C}(U)$ is a non-increasing function from UV to IR as
expected for our special solution. Therefore, the boundary theory
dual to the degenerate aAdS solution does not violate the c-theorem
\cite{Freedman:1999gp} in our case. Its monotonicity has been
checked for the flows considered here.

\section{BPS domain wall}
 In this section, we follow the well-known procedure \cite{Skenderis:1999mm}\cite{Cvetic:1999fe} to discuss the 5D gravity solution with $V_1=0$ and show that it is a BPS domain wall solution. We will introduce
the bosonic differentiable functional by using Euler-Lagrange
equation method to study the stability. In Lagrangian mechanics,
evolution of a physical system is described by the solutions to the
Euler-Lagrange equation (equations of motion
(\ref{generalEOM1})(\ref{generalEOM2})) for the action of the
system. The Euler-Lagrange functional \cite{Skenderis:1999mm} is
 as below \bea\label{energy} E[A_E,\phi] &=&
\int_{0}^\infty\! dz\, {L e^{ A_E(z)}\over z}\Big[{4\over
3}\left({z\over L e^{A_E(z)}}\partial_z\phi\right)^2\nonumber\\
&{}& - 12\Big({z\over L e^{A_E(z)}}\partial_z \left(A_E-\log({z\over
L})\right)\Big)^2 + V_E\Big]\, . \eea We just use the conformal
frame ansatz (\ref{Ametric-Einsteinframe}) to write down the functional
$E$. Note that we have restored the $AdS$ radius $L$. One can differentiate with respect to $A_E$ or $\phi$ to obtain the equation of
motion (\ref{secorder2}) (\ref{secorder3}) as a consistency check.
The function (\ref{energy}) can be rewritten, {\` a} la Bogomol'nyi,
as \bea\label{BPS} E &=&\int_{0}^\infty\! dz \, {L e^{ A_E(z)}\over
z}\left\{ [{z\over L e^{A_E(z)}}\sqrt{4\over 3}\partial_z\phi \mp
3\sqrt{3} W'(\phi)]^2 -12[{z\over L e^{A_E(z)}}\partial_z \left(A_E-\log({z\over L})\right) \pm 2W]^2\right\} \nonumber\\
&& \qquad \pm\, 6[e^{A_E-\log({z\over L})}W]_{0}^\infty\, . \eea
 After introducing an auxiliary field $W(\phi)$ which is called superpotential later, the Einstein-dilaton
equations can be replaced by the following first-order equations. Here
we should require that the following equations hold
  \bea \label{cond}
{z\over L  e^{A_E(z)}}\partial_z \left(A_E-\log({z\over L})\right) &=& \mp 2 W(\phi), \nonumber\\
{z\over L e^{A_E(z)}}\partial_z \phi  &=& \pm {9\over 2}W'(\phi)\, .
\eea We just consider the degenerate solution given by taking $V_1=0$ in
(\ref{solutionexact1}). Luckily we obtain a superpotential analytically from the above formula.
The superpotential $W(\phi)$ is given by
\begin{eqnarray}\label{superpotential}
W\left(\phi\right)={1\over 2 L} \cosh \left(\frac{2 \phi
}{3}\right).
\end{eqnarray}
 The result of \cite{Townsend:1984iu} tells us
that stability constrains the form of the scalar potential to be
\bea\label{dilatonpotential} V_E= 12\left[{9\over 4}(W'(\phi))^2 - 4
W^2\right], \eea where $W(\phi)$ is the superpotential, which is an
arbitrary function and the prime indicates the differentiation with
respect to $\phi$.  As a consistency check, one can plug
(\ref{superpotential}) into (\ref{dilatonpotential}) to obtain
$V_E\left(\phi\right)$. If there exists a real solution of linear
perturbation of $\phi$, it will define a bound of stability named as
the BF bound \cite{Townsend:1984iu}. Therefore the fact that the
dilaton potential can be expressed as the form of
(\ref{dilatonpotential})
 is equivalent to that mass of the scalar field does
not violate the BF bound.

Another possibility to obtain the first order equations (\ref{cond}) is
to calculate the Witten-Nester energy \cite{Witten:1981mf}.
 From the Witten-Nester energy, one can obtain the same superpotential (\ref{superpotential}), as we
will see below.

The Witten-Nester positive energy theorem \cite{Witten:1981mf}
 is
motivated by the fact that the Hamiltonian in supersymmetric
theories is the square of supercharges. This implies that there is
an expression for the energy in terms of spinors and that the energy
is positive definite. The construction below imitates the
supersymmetric argument but does not require supersymmetry. The
spinor formalism of Witten and Nester provides a generalized
``energy'' $E_{WN}$ \cite{Freedman:1999gp}\cite{Gibbons:1982jg}\bea \label{nestE} E_{WN} =
\int_{\partial\Sigma} *\hat
 E,
\eea where $\partial\Sigma$ is the 3D spatial hypersurface in
the whole manifold and $*$ corresponds to the hodge star in 3D
space. The Witten-Nester spinorial energy $E_{WN}$
\cite{Freedman:1999gp}\cite{Gibbons:1982jg}
 is derived using the following
antisymmetric tensor constructed from a spinor fields $\epsilon$
\bea \label{3.nester} \hat E^{\mu\nu} = \bar\epsilon \,
\Gamma^{\mu\nu\rho} \hat\nabla_\rho \epsilon -
\overline{\hat\nabla_\rho\epsilon} \, \Gamma^{\mu\nu\rho} \epsilon
\eea with \bea \label{3.cov} \hat\nabla_\mu = \nabla_\mu + W(\phi)
\Gamma_\mu, \eea where $\nabla$ is the covariant derivative
associated with the metric $g$ and $W(\phi)$ the superpotential.
$\Gamma_{\mu\nu\rho}$ is the 5D antisymmetric gamma matrix. Stokes
theorem states that \bea E_{WN} &=& \int_\Sigma d \Sigma_\mu
\nabla_\nu \hat{E}^{\mu\nu}.\eea Note that this step requires that
the background and deformed solutions are non-singular. The singular
case is beyond the scope of this paper. Readers interested in this issue may refer to \cite{Gibbons:1982jg}. Just imposing the convariant derivative on $E_{WN}$ and making use of the equations of motion
(\ref{generalEOM1}) (\ref{generalEOM2}), one can obtain that
 \bea
\label{hotstuff} \label{3.vol} E_{WN} &=& \int_\Sigma d\Sigma_\mu
\left[ 2 \overline{\delta \psi}_\nu \Gamma^{\mu\nu\rho} \delta
\psi_\rho
- \frac{1}{2} \overline{\delta \chi} \Gamma^\mu \delta \chi \right. \\
&& \left. + \bar{\epsilon} \Gamma^\mu \epsilon \left( -
{V_E(\phi)\over 2}+18 ( {3\over 4} W'(\phi)^2 - \frac{4}{3} W^2 )
\right) \right] \nonumber \eea  with  \bea \label{tino}
\delta \psi_\mu &=& \hat{\nabla}_\mu \epsilon, \\
\delta \chi &=& ({2 \over\sqrt{3}}{\Gamma}^\mu \nabla_\mu \phi -
{3\sqrt{3} } W'(\phi))\epsilon, \label{dil} \eea where
$W'(\phi)=\partial_\phi W$. One can check that (\ref{hotstuff}) is
consistent with the result in \cite{Freedman:2003ax}. Here we do not
calculate the final result step by step, bu just list the
most important formulas leading us to (\ref{hotstuff})(\ref{tino})(\ref{dil}).

In obtaing the above results we have used the following formula \bea \label{epsilon}
\overline{\nabla_\mu \epsilon} \Gamma^{\mu\nu\rho}\nabla_\nu(\omega)
\epsilon = - \overline{ \epsilon}
\Gamma^{\mu\nu\rho}\nabla_\mu\nabla_\nu(\omega) \epsilon, \eea where
$w$ is the spin connection in curved space. The relation
$\nabla_{[\mu}\nabla_{\nu ]}\epsilon= \frac{1}{8} R_{\mu\nu}^{ab}
\Gamma_{ab} \epsilon$ is helpful to realize (\ref{hotstuff}), where $\mu, \nu,\rho$ stand for the indices of
curved space and $a,b$ denote the coordinates of local flat space.
The right hand side of (\ref{epsilon}) can be expanded using the
clifford algebra and gives \bea R_{\mu\nu}^{ab} \bar{\epsilon}
\Gamma^{\mu\nu\rho}\Gamma_{ab}\epsilon &=& R_{\mu\nu}^{ab}
\bar{\epsilon} \{ \Gamma^{\mu\nu\rho}_{ab}+ 6
\delta^{[\rho}_{[a}\delta\Gamma^{\mu\nu]}_{b]}+ 6
\delta^{[\rho}_{[a}\delta^{\nu}_{b]}\Gamma^{\mu]}_{}\}\epsilon \nonumber\\
&=& R_{\mu\nu}^{ab} \bar{\epsilon} \{ \Gamma^{\mu\nu\rho}_{ab}+ 6
\delta^{[\rho}_{[a}\delta\Gamma^{\mu\nu]}_{b]}\}\nonumber\\
& &+ R_{\mu\nu}^{ab} \bar{\epsilon}\{
\delta^{\rho}_{a}\delta^{\nu}_{b}\Gamma^{\mu}-\delta^{\rho}_{a}\delta^{\mu}_{b}\Gamma^{\nu}+\delta^{\mu}_{a}\delta^{\rho}_{b}\Gamma^{\mu}\nonumber\\
& &
-\delta^{\mu}_{a}\delta^{\nu}_{b}\Gamma^{\rho}+\delta^{\nu}_{a}\delta^{\mu}_{b}\Gamma^{\rho}-\delta^{\nu}_{a}\delta^{\rho}_{b}\Gamma^{\mu}\}\epsilon\nonumber\\
&=&  R_{\mu\nu}^{ab} \bar{\epsilon} \{
\Gamma^{\mu\nu\rho}_{ab}\}\epsilon + \bar{\epsilon}\{4
R_{\mu}^{\rho}\Gamma^{\mu}-2 R\Gamma^{\rho}\}\epsilon\nonumber\\
&=&  R_{\mu\nu}^{ab} \bar{\epsilon} \{
\Gamma^{\mu\nu\rho}_{ab}\}\epsilon + 4\{ R^{\mu\rho}-{1\over 2}R
g^{\mu\rho}\}\bar{\epsilon}\Gamma_\mu\epsilon
\label{geteinstein},\eea where we just impose the covariant
derivative on (\ref{3.nester}), and we used
(\ref{geteinstein}) to get the second term in (\ref{3.vol}). One can
make use of Einstein equation to replace $R^{\mu\rho}-{1\over 2}R
g^{\mu\rho}$ appearing in (\ref{geteinstein}) to obtain
(\ref{3.cov}). The first term in ( \ref{geteinstein}) will be
canceled by $ \overline{ \epsilon}
\Gamma^{\mu\nu\rho}\nabla_\mu\hat{\nabla}_\nu(\omega) \epsilon $
automatically. Through straightforward calculation, one can get the
the final form (\ref{3.vol}).
 The last term in
(\ref{3.vol}) will be canceled, if we require that \bea
\label{VWrel} V_E(\phi) =36 \left({3\over 4} W'(\phi)^2 -
\frac{4}{3} W^2(\phi) \right). \eea This constraint is consistent
with the previous statement (\ref{dilatonpotential}). Starting from
(\ref{hotstuff}), one can read out the Witten equation: \bea
\label{bkgdkil}
({\nabla}_\mu +\Gamma_\mu W(\phi))\epsilon &=&0,\\
({\Gamma}^\mu {\nabla}_\mu \phi -{9\over 2} W'(\phi)) \epsilon &=
&0, \eea where $ W(\phi)$ is given by (\ref{superpotential}).
 The spinor solutions of (\ref{bkgdkil}) are the
background Killing spinors similar to those given in
\cite{Skenderis:1999mm} \cite{Freedman:2003ax}: \bea \label{kilspin}
\epsilon &=& e^{\frac{B(r)}{2}} \epsilon_0 = \sqrt{\frac{{L}}{{z}}}
e^{{A_E(z)\over 2}}\epsilon_0, \nonumber
\\
\Gamma^{\hat{r}} \epsilon_0 &=&\epsilon_0 =\Gamma^{\hat{z}}
\epsilon_0, \eea where $\epsilon_0$
is a constant spinor which is chiral with respect to the radial
component of $r$ or $z$. In the last equation we have used the coordinate
transformation $dr =-\frac{L e^{A_E(z)}}{z} dz $.  We confirm that our solution
(\ref{solutionexact1}) with $V_1=0$ is a BPS one. For more complicated
case, the authors of \cite{Cheng:2005wk} have discussed the
asymptotical solution of the killing spinor equation.

\section{Conclusion and discussion}
We apply the potential reconstruction approach to study the graviton
coupled to dilaton system and generate 5D $aAdS$ black hole
solutions in a semi-analytical way. In our approach, there are two
scenarios to obtain the solution. The first is, we choose the
special conformal factor $A_s$ or $A_E$ and then determine the
dilaton configuration $\phi(z)$ from the $tt$ component of the
Einstein equation. The black hole solution $f(z)$ can subsequently
be defined and the corresponding self-interacting potential
$V_s(\phi)$ or $V_E(\phi)$ then determined. In the second scenario,
one chooses a dilation configuration $\phi(z)$ to define the
conformal factor $A_s$ or $A_E$. The remaining steps of
reconstructing the gravity background are the same as in the first
scenario. We study the related properties of a simple 5D gravity
solution (\ref{solutionexact1}) in the remaining part.

We also investigate the hydrodynamics of the boundary fluid dual to
our bulk gravitational solution, following the spirit of the
fluid/gravity correspondence. We worked out the asymptotic behavior
of the correction terms that need to be added to achieve an exact
solution of the bulk equation of motion after promoting the
parameters of the original solution to slowly-varying functions of
the boundary coordinates. In particular, we presented the asymptotic
form of the $SO(3)$ tensor components of the correction terms, which
is useful for extracting the ratio of the shear viscosity to the
entropy density of the boundary fluid. Another interesting observation is that, the dual boundary fluid is conformally anomalous, and the anomaly is related to the temperature of the black hole. This might be a potential useful feature of our solution in describing physical situations where conformal symmetry is broken.

Furthermore, we study the holographic conformal anomaly
characterized by the c function and the stability of the degenerate
aAdS solution as a special case of our 5D black hole solution. We
check that the degenerate solution does not violate the c theorem.
The dual RG flow of the holographic conformal anomaly owns the
monotonicity and this situation is a special case of
\cite{Skenderis:1999mm}. Following \cite{Skenderis:1999mm}
\cite{Cvetic:1999fe}, we list the formalism for the bosonic
functional energy and the Witten-Nester energy in conformal
coordinates. We check that the degenerate aAdS solution is a BPS
domain wall solution.

For future studies, it is valuable and feasible to use the potential
reconstruction approach to reconstruct the gravity solutions
revelent for phenomenology. For example, to describe more realistic
case, one can impose constraints on the metric from experimental
data and use the method to reconstruct the gravity system
corresponding to the metric. It is an efficient way to link the
gravity and phenomena in field theory directly. It will apply to
various fields, such as: AdS/QCD, AdS/CMT, fluid/gravity dualtiy and
so forth. It opens a new window to test the gauge/gravity duality.

 \vskip 1cm \noindent {\bf
Acknowledgments}: The authors thank Rong-Gen Cai, Mei Huang, Bin Hu,
Jen-Chi Lee, Hong Lu, Danning Li, Li Li, JunBao Wu, QiShu Yan, Yi
Yang, YunLong Zhang, HaiQing Zhang for valuable discussions. S.H.
appreciates the hospitality of the institute of high energy physics,
CAS during the initial stage of this work. Y.P Hu is supported by
China Postdoctoral Science Foundation under Grant No.20110490227 and
National Natural Science Foundation of China (NSFC) under grant
No.11105004. This work is also supported partially by grants from
NSFC, No. 10975168 and No. 11035008.


\begin{thebibliography}{99}
\bibitem{Maldacena:1997re}
  J.~M.~Maldacena,
  ``The large N limit of superconformal field theories and supergravity,''
  Adv.\ Theor.\ Math.\ Phys.\  {\bf 2}, 231 (1998)
  [Int.\ J.\ Theor.\ Phys.\  {\bf 38}, 1113 (1999)]  [arXiv:hep-th/9711200].
\bibitem{Gubser:1998bc}
  S.~S.~Gubser, I.~R.~Klebanov and A.~M.~Polyakov,
  ``Gauge theory correlators from non-critical string theory,''
  Phys.\ Lett.\  B {\bf 428}, 105 (1998)
  [arXiv:hep-th/9802109].
\bibitem{Witten:1998qj}
  E.~Witten,
  ``Anti-de Sitter space and holography,''
  Adv.\ Theor.\ Math.\ Phys.\  {\bf 2}, 253 (1998)
  [arXiv:hep-th/9802150].
\bibitem{Aharony:1999ti}
  O.~Aharony, S.~S.~Gubser, J.~M.~Maldacena, H.~Ooguri and Y.~Oz,
  ``Large N field theories, string theory and gravity,''
  Phys.\ Rept.\  {\bf 323}, 183 (2000)
  [arXiv:hep-th/9905111].

\bibitem{Martinez:2004nb}
  C.~Martinez, R.~Troncoso, J.~Zanelli,
  ``Exact black hole solution with a minimally coupled scalar field,''
  Phys.\ Rev.\  {\bf D70}, 084035 (2004).
  [hep-th/0406111].

\bibitem{Dotti:2007cp}
  G.~Dotti, R.~J.~Gleiser, C.~Martinez,
  ``Static black hole solutions with a self interacting conformally coupled scalar field,''
  Phys.\ Rev.\  {\bf D77}, 104035 (2008).
  [arXiv:0710.1735 [hep-th]].




\bibitem{Hatsuda:2002hz}
  M.~Hatsuda, M.~Sakaguchi,
  ``Wess-Zumino term for AdS superstring,''
  Phys.\ Rev.\  {\bf D66}, 045020 (2002).
  [hep-th/0205092].

\bibitem{Zeng:2009fp}
  D.~-f.~Zeng,
  ``An Exact Hairy Black Hole Solution for AdS/CFT Superconductors,''
  [arXiv:0903.2620 [hep-th]].

\bibitem{Torii:2001pg}
  T.~Torii, K.~Maeda, M.~Narita,
  ``Scalar hair on the black hole in asymptotically anti-de Sitter space-time,''
  Phys.\ Rev.\  {\bf D64}, 044007 (2001).


\bibitem{Megias:2010ku}
  E.~Megias, H.~J.~Pirner and K.~Veschgini,
  ``QCD-Thermodynamics using 5-dim Gravity,''
  Phys.\ Rev.\  D {\bf 83}, 056003 (2011)
  [arXiv:1009.2953 [hep-ph]];
  K.~Veschgini, E.~Megias and H.~J.~Pirner,
  ``Trouble Finding the Optimal AdS/QCD,''
  Phys.\ Lett.\  B {\bf 696}, 495 (2011)
  [arXiv:1009.4639 [hep-th]];
  E.~Megias, H.~J.~Pirner and K.~Veschgini,
  ``Thermodynamics of AdS/QCD within the 5D dilaton-gravity model,''
  Nucl.\ Phys.\ Proc.\ Suppl.\  {\bf 207-208}, 333 (2010)
  [arXiv:1008.4505 [hep-th]];
  B.~Galow, E.~Megias, J.~Nian and H.~J.~Pirner,
  ``Phenomenology of AdS/QCD and Its Gravity Dual,''
  Nucl.\ Phys.\  B {\bf 834}, 330 (2010)
  [arXiv:0911.0627 [hep-ph]].



\bibitem{Gubser-T}
S.~S.~Gubser and A.~Nellore,
  ``Mimicking the QCD equation of state with a dual black hole,''
  Phys.\ Rev.\  D {\bf 78}, 086007 (2008);
S.~S.~Gubser, A.~Nellore, S.~S.~Pufu and F.~D.~Rocha,
``Thermodynamics and bulk viscosity of approximate black hole duals
to finite temperature quantum chromodynamics,''
  Phys.\ Rev.\ Lett.\  {\bf 101}, 131601 (2008);
  S.~S.~Gubser, S.~S.~Pufu and F.~D.~Rocha,
  ``Bulk viscosity of strongly coupled plasmas with holographic duals,''
  JHEP {\bf 0808}, 085 (2008).

\bibitem{Gursoy-T}
U.~Gursoy, E.~Kiritsis, L.~Mazzanti and F.~Nitti, ``Deconfinement
and Gluon Plasma Dynamics in Improved Holographic QCD,'' Phys.\
Rev.\ Lett.\  {\bf 101}, 181601 (2008);
U.~Gursoy, E.~Kiritsis, G.~Michalogiorgakis and F.~Nitti, ``Thermal
Transport and Drag Force in Improved Holographic QCD,''
  JHEP {\bf 0912}, 056 (2009).

\bibitem{Behrndt:1998jd}
  K.~Behrndt, M.~Cvetic, W.~A.~Sabra,
  ``Nonextreme black holes of five-dimensional N=2 AdS supergravity,''
  Nucl.\ Phys.\  {\bf B553}, 317-332 (1999).
  [hep-th/9810227].
\bibitem{Cai:1998ji}
  R.~-G.~Cai, K.~-S.~Soh,
  ``Critical behavior in the rotating D-branes,''
  Mod.\ Phys.\ Lett.\  {\bf A14}, 1895-1908 (1999).
  [hep-th/9812121].

\bibitem{Gubser:1998jb}
  S.~S.~Gubser,
  ``Thermodynamics of spinning D3-branes,''
  Nucl.\ Phys.\  {\bf B551}, 667-684 (1999).
  [hep-th/9810225].
\bibitem{Benincasa:2005iv}
  P.~Benincasa, A.~Buchel, A.~O.~Starinets,
  ``Sound waves in strongly coupled non-conformal gauge theory plasma,''
  Nucl.\ Phys.\  {\bf B733}, 160-187 (2006).
  [hep-th/0507026].

\bibitem{Benincasa:2006ei}
  P.~Benincasa, A.~Buchel,
  ``Hydrodynamics of Sakai-Sugimoto model in the quenched approximation,''
  Phys.\ Lett.\  {\bf B640}, 108-115 (2006).
  [hep-th/0605076].


\bibitem{Sakai:2004cn}
  T.~Sakai, S.~Sugimoto,
  ``Low energy hadron physics in holographic QCD,''
  Prog.\ Theor.\ Phys.\  {\bf 113}, 843-882 (2005).
  [arXiv:hep-th/0412141 [hep-th]].

\bibitem{ChiouLahanas:2009cs}
  C.~Chiou-Lahanas, G.~A.~Diamandis and B.~C.~Georgalas,
  ``Five-Dimensional Black Hole String Backgrounds and Brane Universe
  Phys.\ Lett.\  B {\bf 678}, 485 (2009)
  [arXiv:0904.1484 [hep-th]].





\bibitem{Farakos:2009fx}
  K.~Farakos, A.~P.~Kouretsis, P.~Pasipoularides,
  ``Anti de Sitter 5D black hole solutions with a self-interacting bulk scalar field: A Potential reconstruction approach,''
  Phys.\ Rev.\  {\bf D80}, 064020 (2009).
  [arXiv:0905.1345 [hep-th]].

\bibitem{Li:2011hp}
  D.~Li, S.~He, M.~Huang and Q.~S.~Yan,
  ``Thermodynamics of deformed AdS$_5$ model with a positive/negative quadratic
  correction in graviton-dilaton system,''
  JHEP {\bf 1109}, 041 (2011)
  [arXiv:1103.5389 [hep-th]].

\bibitem{He:2010ye}
  S.~He, M.~Huang, Q.~-S.~Yan,
  ``Logarithmic correction in the deformed $AdS_5$ model to produce the heavy quark potential and QCD beta function,''
  Phys.\ Rev.\  {\bf D83}, 045034 (2011).
  [arXiv:1004.1880 [hep-ph]].


\bibitem{Ohta:2009pe}
  N.~Ohta, T.~Torii,
  ``Black Holes in the Dilatonic Einstein-Gauss-Bonnet Theory in Various Dimensions IV: Topological Black Holes with and without Cosmological Term,''
  Prog.\ Theor.\ Phys.\  {\bf 122}, 1477-1500 (2009).
  [arXiv:0908.3918 [hep-th]].

\bibitem{Kolyvaris:2009pc}
  T.~Kolyvaris, G.~Koutsoumbas, E.~Papantonopoulos, G.~Siopsis,
  ``A New Class of Exact Hairy Black Hole Solutions,''
  Gen.\ Rel.\ Grav.\  {\bf 43}, 163-180 (2011).
  [arXiv:0911.1711 [hep-th]].






\bibitem{Brown}
  J.~D.~Brown and J.~W.~York,
  ``Quasilocal energy and conserved charges derived from the gravitational
  action,''
  Phys.\ Rev.\  D {\bf 47}, 1407 (1993)
  [arXiv:gr-qc/9209012];
  J.~D.~Brown, J.~Creighton and R.~B.~Mann,
  ``Temperature, energy and heat capacity of asymptotically anti-de Sitter
  black holes,''
  Phys.\ Rev.\  D {\bf 50}, 6394 (1994)
  [arXiv:gr-qc/9405007].




\bibitem{Counterterm}
  M.~Henningson and K.~Skenderis,
  ``The holographic Weyl anomaly,''
  JHEP {\bf 9807}, 023 (1998)
  [arXiv:hep-th/9806087];
  V.~Balasubramanian and P.~Kraus,
  ``A stress tensor for anti-de Sitter gravity,''
  Commun.\ Math.\ Phys.\  {\bf 208}, 413 (1999)
  [arXiv:hep-th/9902121];
  K.~Skenderis,
  ``Lecture notes on holographic renormalization,''
  Class.\ Quant.\ Grav.\  {\bf 19}, 5849 (2002)
  [arXiv:hep-th/0209067];
  R.~C.~Myers,
  ``Stress tensors and Casimir energies in the AdS/CFT correspondence,''
  Phys.\ Rev.\  D {\bf 60}, 046002 (1999)
  [arXiv:hep-th/9903203];
  S.~de Haro, S.~N.~Solodukhin and K.~Skenderis,
  Commun.\ Math.\ Phys.\  {\bf 217}, 595 (2001)
  [arXiv:hep-th/0002230].


\bibitem{Bhattacharyya:2008jc}
  S.~Bhattacharyya, V.~E.~Hubeny, S.~Minwalla and M.~Rangamani,
  ``Nonlinear Fluid Dynamics from Gravity,''
  JHEP {\bf 0802}, 045 (2008)
  [arXiv:0712.2456 [hep-th]];

\bibitem{Iqbal:2008by}
  N.~Iqbal and H.~Liu,
  ``Universality of the hydrodynamic limit in AdS/CFT and the membrane
  paradigm,''
  arXiv:0809.3808 [hep-th].

\bibitem{Cai:2008ph}
  R.~G.~Cai, Z.~Y.~Nie and Y.~W.~Sun,
  ``Shear Viscosity from Effective Couplings of Gravitons,''
  arXiv:0811.1665 [hep-th];
  R.~G.~Cai, Z.~Y.~Nie, N.~Ohta and Y.~W.~Sun,
  ``Shear Viscosity from Gauss-Bonnet Gravity with a Dilaton Coupling,''
  Phys.\ Rev.\  D {\bf 79}, 066004 (2009)
  [arXiv:0901.1421 [hep-th]].

\bibitem{Policastro:2001yc}
  G.~Policastro, D.~T.~Son and A.~O.~Starinets,
  ``The shear viscosity of strongly coupled N = 4 supersymmetric Yang-Mills
  plasma,''
  Phys.\ Rev.\ Lett.\  {\bf 87}, 081601 (2001)
  [arXiv:hep-th/0104066].
\bibitem{Kovtun:2003wp}
  P.~Kovtun, D.~T.~Son and A.~O.~Starinets,
  ``Holography and hydrodynamics: Diffusion on stretched horizons,''
  JHEP {\bf 0310}, 064 (2003)
  [arXiv:hep-th/0309213].
\bibitem{Buchel:2003tz}
  A.~Buchel and J.~T.~Liu,
  ``Universality of the shear viscosity in supergravity,''
  Phys.\ Rev.\ Lett.\  {\bf 93}, 090602 (2004)
  [arXiv:hep-th/0311175].
\bibitem{Kovtun:2004de}
  P.~Kovtun, D.~T.~Son and A.~O.~Starinets,
  ``Viscosity in strongly interacting quantum field theories from black hole
  physics,''
  Phys.\ Rev.\ Lett.\  {\bf 94}, 111601 (2005)
  [arXiv:hep-th/0405231].



\bibitem{Mas:2006dy}
  J.~Mas,
  ``Shear viscosity from R-charged AdS black holes,''
  JHEP {\bf 0603}, 016 (2006)
  [arXiv:hep-th/0601144].
\bibitem{Son:2006em}
  D.~T.~Son and A.~O.~Starinets,
  ``Hydrodynamics of R-charged black holes,''
  JHEP {\bf 0603}, 052 (2006)
  [arXiv:hep-th/0601157].
\bibitem{Saremi:2006ep}
  O.~Saremi,
  ``The viscosity bound conjecture and hydrodynamics of M2-brane theory at
  finite chemical potential,''
  JHEP {\bf 0610}, 083 (2006)
  [arXiv:hep-th/0601159].
\bibitem{Maeda:2006by}
  K.~Maeda, M.~Natsuume and T.~Okamura,
  ``Viscosity of gauge theory plasma with a chemical potential from  AdS/CFT,''
  Phys.\ Rev.\  D {\bf 73}, 066013 (2006)
  [arXiv:hep-th/0602010].
\bibitem{Cai:2008in}
  R.~G.~Cai and Y.~W.~Sun,
  ``Shear Viscosity from AdS Born-Infeld Black Holes,''
  JHEP {\bf 0809}, 115 (2008)
  [arXiv:0807.2377 [hep-th]].

\bibitem{Sinha}
  A.~Buchel, R.~C.~Myers, M.~F.~Paulos and A.~Sinha,
  ``Universal holographic hydrodynamics at finite coupling,''
  Phys.\ Lett.\  B {\bf 669}, 364 (2008)
  [arXiv:0808.1837 [hep-th]];
\bibitem{Myers:2008yi}
  R.~C.~Myers, M.~F.~Paulos and A.~Sinha,
  Phys.\ Rev.\  D {\bf 79}, 041901 (2009)
  [arXiv:0806.2156 [hep-th]].



\bibitem{Alvarez:1998wr}
  E.~Alvarez, C.~Gomez,
  ``Geometric holography, the renormalization group and the c theorem,''
  Nucl.\ Phys.\  {\bf B541}, 441-460 (1999).
  [hep-th/9807226].

\bibitem{Henningson:1998gx}
  M.~Henningson and K.~Skenderis,
  ``The Holographic Weyl anomaly,''
  JHEP {\bf 9807}, 023 (1998)
  [arXiv:hep-th/9806087].
  M.~Henningson, K.~Skenderis,
  ``Holography and the Weyl anomaly,''
  Fortsch.\ Phys.\  {\bf 48}, 125-128 (2000).
  [hep-th/9812032].


\bibitem{Distler:1998gb}
  J.~Distler, F.~Zamora,
  ``Nonsupersymmetric conformal field theories from stable anti-de Sitter spaces,''
  Adv.\ Theor.\ Math.\ Phys.\  {\bf 2}, 1405-1439 (1999).
  [hep-th/9810206].

\bibitem{Freedman:1999gp}
  D.~Z.~Freedman, S.~S.~Gubser, K.~Pilch, N.~P.~Warner,
  ``Renormalization group flows from holography supersymmetry and a c theorem,''
  Adv.\ Theor.\ Math.\ Phys.\  {\bf 3}, 363-417 (1999).
  [hep-th/9904017].

\bibitem{Skenderis:1999mm}
  K.~Skenderis, P.~K.~Townsend,
  ``Gravitational stability and renormalization group flow,''
  Phys.\ Lett.\  {\bf B468}, 46-51 (1999).
  [hep-th/9909070].


\bibitem{Witten:1981mf}
  E.~Witten,
  ``A Simple Proof of the Positive Energy Theorem,''
  Commun.\ Math.\ Phys.\  {\bf 80}, 381 (1981).
  J. Nester, A new gravitational expression with a simple positivity
proof, Phys. Lett. 83A, 241 (1981).



\bibitem{entropy-BK}
  J.~D.~Bekenstein,
  ``Black holes and entropy,''
  Phys.\ Rev.\  D {\bf 7}, 2333 (1973);
 S.~W.~Hawking,
  ``Particle Creation By Black Holes,''
  Commun.\ Math.\ Phys.\  {\bf 43}, 199 (1975)
  [Erratum-ibid.\  {\bf 46}, 206 (1976)].


\bibitem{Hu:2009rj}
  Y.~P.~Hu,
  ``Tension term, interchange symmetry, and the analogy of energy and tension
  JHEP {\bf 0905}, 096 (2009)
  [arXiv:0904.1250 [hep-th]].




\bibitem{Nojiri:1998dh}
  S.~Nojiri and S.~D.~Odintsov,
  ``Conformal anomaly for dilaton coupled theories from AdS/CFT
  correspondence,''
  Phys.\ Lett.\  B {\bf 444}, 92 (1998)
  [arXiv:hep-th/9810008].



\bibitem{Cai:1999xg}
  R.~G.~Cai and N.~Ohta,
  ``Surface counterterms and boundary stress-energy tensors for  asymptotically
  non-anti-de Sitter spaces,''
  Phys.\ Rev.\  D {\bf 62}, 024006 (2000)
  [arXiv:hep-th/9912013].

\bibitem{Nojiri:2000kh}
  S.~Nojiri, S.~D.~Odintsov and S.~Ogushi,
  ``Finite action in d5 gauged supergravity and dilatonic conformal anomaly
  for dual quantum field theory,''
  Phys.\ Rev.\  D {\bf 62}, 124002 (2000)
  [arXiv:hep-th/0001122].




\bibitem{Hur:2008tq}
  J.~Hur, K.~K.~Kim and S.~J.~Sin,
  ``Hydrodynamics with conserved current from the gravity dual,''
  JHEP {\bf 0903}, 036 (2009)
  [arXiv:0809.4541 [hep-th]].

\bibitem{Hu:2011ze}
  Y.~-P.~Hu, P.~Sun, J.~-H.~Zhang,
  ``Hydrodynamics with conserved current via AdS/CFT correspondence in the Maxwell-Gauss-Bonnet gravity,''
  Phys.\ Rev.\ {\bf D83 } (2011)  126003.
  [arXiv:1103.3773 [hep-th]];  Y.~-P.~Hu, H.~-F.~Li, Z.~-Y.~Nie,
  ``The first order hydrodynamics via AdS/CFT correspondence in the Gauss-Bonnet gravity,''
  JHEP {\bf 1101}, 123 (2011).
  [arXiv:1012.0174 [hep-th]].

\bibitem{Myers:1999psa}
  R.~C.~Myers,
  ``Stress tensors and Casimir energies in the AdS / CFT correspondence,''
  Phys.\ Rev.\  {\bf D60}, 046002 (1999).
  [hep-th/9903203].




\bibitem{Cvetic:1999fe}
  M.~Cvetic, H.~Lu and C.~N.~Pope,
  ``Domain walls and massive gauged supergravity potentials,''
  Class.\ Quant.\ Grav.\  {\bf 17}, 4867 (2000)
  [arXiv:hep-th/0001002].




\bibitem{Townsend:1984iu}
  P.~K.~Townsend,
  ``Positive energy and the scalar potential in higher dimensional super gravity theores,''
  Phys.\ Lett.\  B {\bf 148}, 55 (1984).



\bibitem{Gibbons:1982jg}
  G.~W.~Gibbons, S.~W.~Hawking, G.~T.~Horowitz, M.~J.~Perry,
  ``Positive Mass Theorems For Black Holes,''
  Commun.\ Math.\ Phys.\  {\bf 88}, 295 (1983).


\bibitem{Freedman:2003ax}
  D.~Z.~Freedman, C.~Nunez, M.~Schnabl, K.~Skenderis,
  ``Fake supergravity and domain wall stability,''
  Phys.\ Rev.\  {\bf D69}, 104027 (2004).
  [hep-th/0312055].

\bibitem{Cheng:2005wk}
  M.~C.~N.~Cheng, K.~Skenderis,
  ``Positivity of energy for asymptotically locally AdS spacetimes,''
  JHEP {\bf 0508}, 107 (2005).
  [hep-th/0506123].



















\end{thebibliography}
\end{document}